\documentclass{jpsj3}
\usepackage{txfonts}
\usepackage{bm}
\usepackage{amsfonts}
\usepackage{mathrsfs}
\usepackage{amsmath}
\usepackage{color}

\title{Pressure-Driven Topological Phase Transition in the Yb Chalcogenides YbO and YbS}

\author{Zhi Li$^{1,2}$\thanks{E-mail:zhili@hfut.edu.cn} and Jiu-Xing Zhang$^1$\thanks{E-mail:zjiuxing@hfut.edu.cn}}
\inst{$^1$School of Materials Science and Engineering, Hefei University of Technology, Hefei 230009, Anhui, China \\
$^2$Department of Physics, California State University Northridge, Northridge, CA 91330-8268, USA \\
} 

\abst{By first-principles calculation based on the density functional theory (DFT) with the modified Becke-Johnson local density approximation plus Hubbard \emph{U} (MBJLDA+\emph{U}), we studied the band structures of the Yb chalcogenides YbO and YbS under ambient and high pressures. It was revealed that both YbO and YbS have a trivial band topology under ambient pressure, and a nontrivial band topology under high pressure. The topological phase transition is reduced by the pressure-driven single-band inversion between \emph{5d-} and \emph{4f}-orbitals at the time-reversal invariant momentum (TRIM) point X. A bulk Dirac cone coexisting with a pair of metallic surface states on the [001] surface determined by tight binding model calculation with a slab geometry also demonstrates the nontrivial band topology of YbO and YbS under high pressure.}

\begin{document}
\maketitle

\section{Introduction}
Because of the emergent novel surface state protected by nontrivial band topology, research on topological classification for gapped systems is surging, \cite{Ryo10,fang12,Chen13} In the family of symmetry-protected topological order, the topological insulator (TI) protected by time-reversal symmetry attracts a lots of research interest on fundamental theory and realization in concrete materials for potential industrial applications.\cite{Hasan10,XLQ11,Ando13,Moore07} Not only conventional semiconductor materials with strong spin-orbital coupling (SOC),\cite{Fu06} but also Kondo insulator materials are attracting increasing research interest. \cite{Coleman10,Maxim12,Tran12,Coleman13,Werner13,Legner13} In 2010, on the basis of the periodic Anderson Kondo lattice model, Dzero et al. proposed the topological classification in a simple cubic Kondo insulator. \cite{Coleman10} Their work showed that, with the fluctuating valence of a rare-earth element, the transition between topological trivial and nontrivial states can be realized in a simple cubic Kondo insulator prototyped using SmB$_{6}$.\cite{Coleman13}  Both first-principles calculation and experiments suggest that the simple cubic SmB$_{6}$ with a mixed valence state is a candidate topological Kondo insulator (TKI). \cite{DAI13,Neupane13,Jiang13,KIM12,KIM13} PuB$_{6}$ is also predicted as candidate TKI by first-principles calculation. \cite{Deng13}

Very recently, Weng et al. have predicted that the simple cubic YbB$_{6}$ is a candidate topological insulator, while YbB$_{12}$ with a rock-salt structure is a potential candidate topological crystalline insulator with a nontrivial mirror Chern number protected by mirror symmetry.  \cite{DAI14,Saso03} Angle-resolved photoemission spectroscopy (ARPES) also suggested YbB$_{6}$ as a candidate TKI.\cite{Feng14} In fact, there are hundreds of rare-earth compounds RX with a rock-salt structure, with R referring to the rare-earth Ce, Pr, Nd, Pm, Sm, Eu, Gd, Tb, Dy, Ho, Er, Tm, Yb, and Lu, and X standing for the pnictide atoms N, P, As, Sb, Bi and the chalcogenide atoms O, S, Se, Te, Po.\cite{petit10} There are many semiconductors or semimetals with different combinations of R and X. In particular, for the chalcogenides of Eu, Sm, and Yb, \cite{Wills99} a fluctuating valence state is accessible under ambient or high pressure. \cite{Delley90,Svane01,Bucher71,werner81,Jarrige13,Leger85} These rare-earth compounds providing a high probability of finding more TKIs, although there are few TKIs predicted in rare-earth pnictides and chalcogenides with a rock-salt structure at present. \cite{Zhi14}

In this work, we employ first-principles calculation based on the density functional theory (DFT) with the modified Becke-Johnson local density approximation plus Hubbard \emph{U} (MBJLDA+\emph{U}) to study the band topology of the Yb monochalcogenides YbO and YbS with a rock-salt structure (shown in Fig. 1). Experimentally, both YbO and YbS are insulators with band gaps of 0.32 and 1.40 eV respectively, at ambient pressure, \cite{werner81,Jarrige13} and YbO is on the margin of a mixed valence state.\cite{petit10} Our MBJLDA+$U$ calculations predict that, at ambient pressure, both YbO and YbS are trivial insulators with band gaps of about 0.24 and 1.26 eV, respectively, and both of them will undergo a topological phase transition to a nontrivial topological metallic state under high pressure. The insulator-metal transition under high pressure is consistent with the experimental result. \cite{werner81,Jarrige13} The novelty of this work is that we also predict that this transition is a topological phase transition, i.e., the insulating and metallic phases have different band topologies. The mechanism for this topological phase transition is pressure-driven band inversion at the time-reversal invariant momentum (TRIM) point X between the 4\emph{f}-dominating band and  the 5\emph{d} conduction band owing to the increasing bandwidth of 5\emph{d}-derived states under high pressure. At ambient pressure, the Yb atom in both YbO and YbS is divalent because of the electronic correlation mimicked by Hubbard \emph{U}, which localizes the \emph{4f}-orbitals and suppresses the charge transfer between the 4\emph{f-} and 5\emph{d}-orbitals. The suppressed charge transfer between the \emph{4f-} and \emph{5d}-orbitals renders the band topology of YbO trivial. Under high pressure, the bandwidth of the \emph{5d}-orbital clearly increases. Once the band minimum of the \emph{5d}-orbital is lower than the band maximum of the \emph{4f}-orbital, charge transfer between the \emph{4f-} and \emph{5d}-orbitals will take place. The pressure-induced charge transfer will lead to a mixed valence state in YbO and YbS, which is a necessary condition for TKI. \cite{DAI13,DAI14}

\section{Methods}

 The DFT calculations employed the all-electron, full-potential linearized augmented plane wave (FPLAPW) method with MBJLDA implemented in the WIEN2K code.\cite{Wien2k,Tran2009}
To account for the on-site 4$f$-electron correlations, we applied the MBJLDA + \emph{U}  scheme that incorporates the on-site Coulomb repulsion \emph{U} and the Hund's rule coupling strength \emph{J$_H$} with an approximation correction for self-interaction correction (SIC)  for the Yb 4\emph{f}-orbitals.\cite{SIC} The spin-orbit coupling was included in the self-consistent calculations, and the on-site Hund's exchange parameter J$_H$ was set to zero. We used \emph{R$_{MT}$}$\times$\emph{K$_{max}$}=9.0, muffin-tin radii of 2.50 a.u. for Yb, and 2.33 (2.05) a.u. for S (O), and a 30$\times$30$\times$30 k-point Monkhorst-Pack mesh. We used the experimental lattice parameters \textit{a}=4.86 and 5.68 {\AA} for the lattice parameters of YbO and YbS under ambient pressure. Under high pressure, i.e. reduced lattice, we present the calculated band structures with \textit{a}=4.70 and 5.10 {\AA} for YbO and YbS, respectively, where the \emph{d}-band will inverse with the top \emph{f}-band.

The basis set consisted of the Yb 4$f$, 5$d$, 6$s$, and 6$p$ valence states, the Yb 5$s$ and 5$p$ semicore states (treated in another energy window), and the
O, S 3$s$ semicore and 3$p$ valence states.

\begin{figure}[t]
\includegraphics[width=15cm]{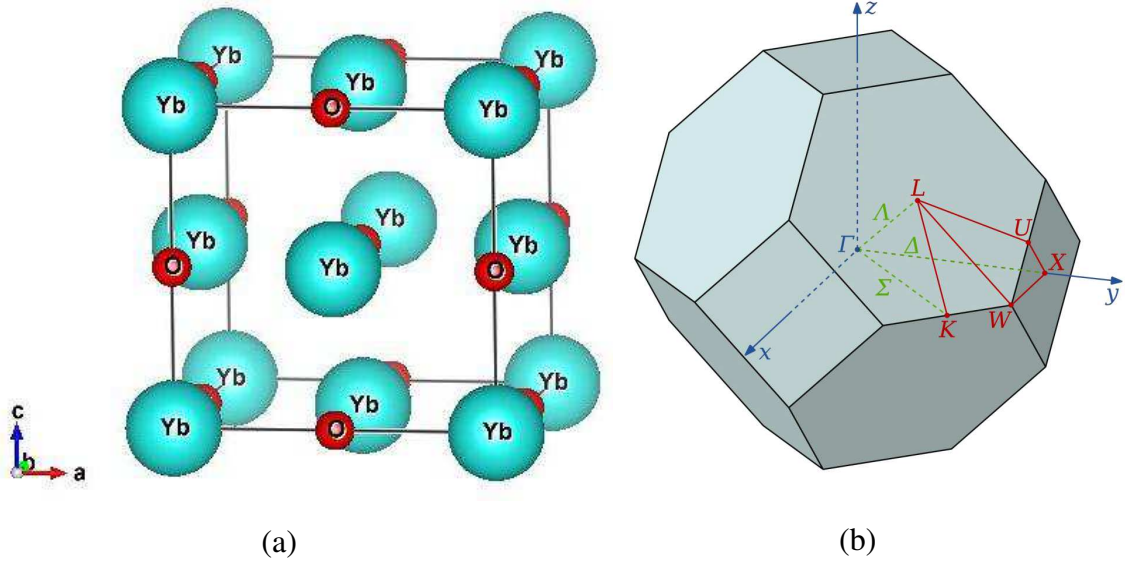}
\caption{\label{fig1} (Color online) Crystal structure of YbO and its Brillouin zone (BZ).}
\end{figure}

\section{Results and Discussion}
\textit{Ambient pressure} The calculated band structures of YbO and YbS at ambient pressure are shown in Fig. 2. Both YbO and YbS are predicted as good metals by MBJLDA calculation, and the band inversion between 5\emph{d-} and 4\emph{f}-orbitals exists in both band structures. However, YbO is a semiconductor with a 0.32 eV band gap, and YbS is an insulator with a 1.40 eV band gap as determined by experiment.\cite{werner81,Jarrige13} While the MBJLDA calculation is crucial for the noninteracting topological insulators,\cite{Yao10} and describes accurately the correlation effects of the 5$d$ states, it can't describe well those of the more localized 4$f$ states, whose treatment still requires an additional Hubbard $U$ value.\cite{Blaha11,Yiang2013}
\begin{figure}[t]
\includegraphics[width=15cm]{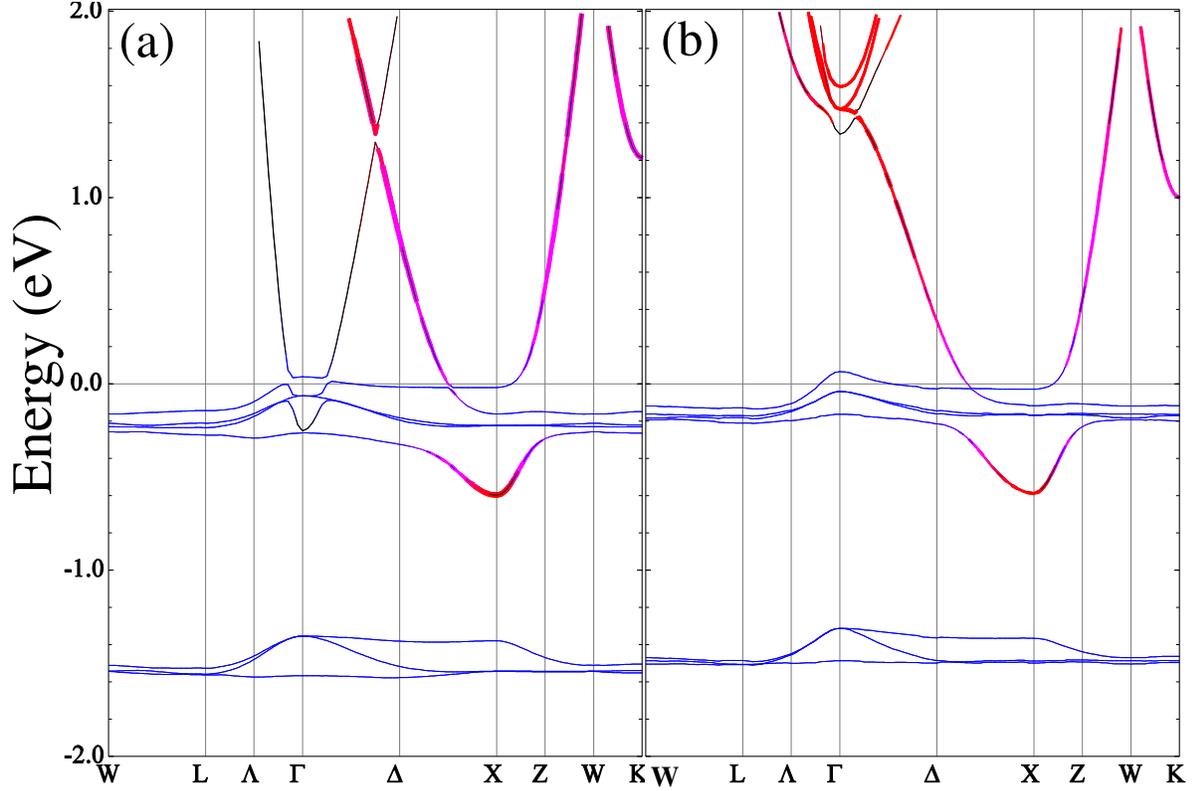}
\caption{\label{fig2} (Color online) Band structures of YbO with experimental lattice parameters a=4.86 {\AA}(a) and YbS with experimental lattice parameter a=5.68 {\AA}(b) obtained by MBJLDA calculation. The weights of 5\emph{d} and 4\emph{f} electrons are represented by dark and gray, respectively.}
\end{figure}

The MBJLDA+$U$ calculations with $U$ of $\sim$ 3.0 eV render a band gap of $\sim$0.24 eV for YbO, while YbS needs a larger Hubbard $U$=7.0 eV to obtain a band gap of 1.26 eV. Our band gap determined by MBJLDA+$U$ calculation is close to the experimental gaps of 0.30 and 1.40 eV for YbO and YbS, respectively.\cite{werner81,Jarrige13} A larger \emph{U} is required to obtain the experimental gap for YbS, whose lattice parameter is larger than that of YbO. This variable \emph{U} is consistent with previous work on Yb, \cite{Pickett09} where a decreasing \emph{U} is required under increasing pressure, i.e., reduced lattice parameter.  The band structures of YbO and YbS determined by MBJLDA+$U$ calculation are shown in Figs. 3(a) and 4(a), respectively, in which SOC split \emph{f}-bands into two sub-bands consisted of \emph{f}$_{5/2}$ and \emph{f}$_{7/2}$ states, respectively. At ambient pressure, \emph{d}- and \emph{f}-bands are well separated and there is no hybridization between \emph{d}- and \emph{f}-orbitals. Because of the empty \emph{d}-band, the Yb ion is divalent. The band topology is trivial since there is no band inversion in the entire Brillouin zone (BZ).
\begin{figure}[t]
\includegraphics[width=15cm]{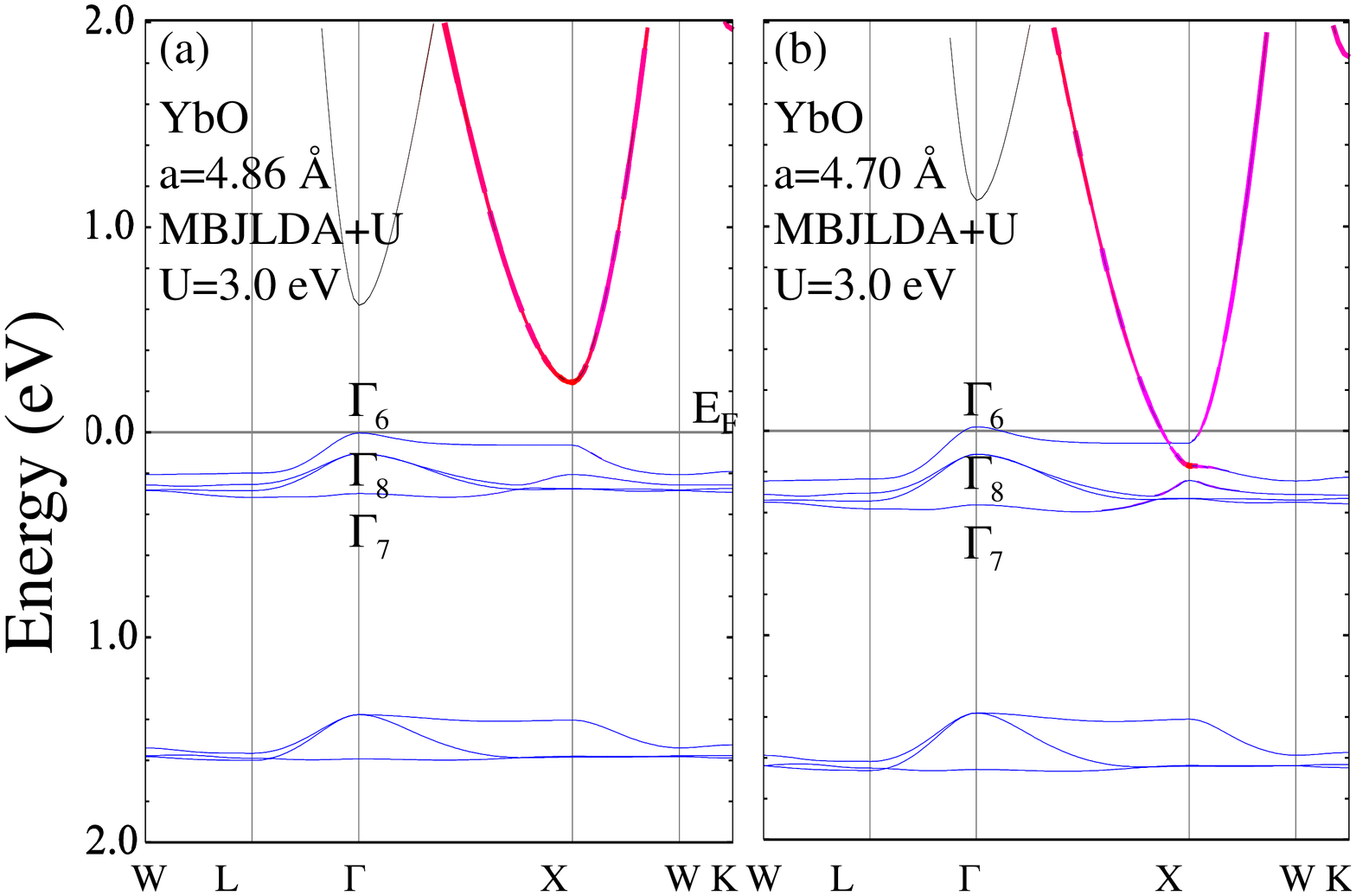}
\caption{\label{fig3} (Color online) Band structure of YbO with experimental lattice parameter a=4.86 (a) and 4.70 {\AA}(b) obtained by MBJLDA+\emph{U} calculation, with \emph{U}=3.0 eV. The weights of 5\emph{d} and 4\emph{f} electrons are represented by dark and gray, respectively.}
\end{figure}

 \textit{High pressure} Even though both YbO and YbS are regular band insulators at ambient pressure, we can still induce a topological phase transition in both of them by applying external pressure. Since there is no band inversion in both YbO and YbS because of the absence of \emph{d}-\emph{f} hybridization and valence fluctuation at ambient pressure, we can apply high pressure approach to extend the wave functions of \emph{5d}- and \emph{4f}-orbitals and realize the charge transfer between \emph{5d-} and \emph{4f}-orbitals.  Once the band minimum of the conducting \emph{5d} band is below the band maximum of the \emph{f}$_{7/2}$ band, charge transfer between \emph{5d-} and \emph{4f}-orbitals will take place, which will cause the system to be in a mixed valence state. The calculated band structures with the reduced lattice parameters \textit{a}=4.70 {\AA} for YbO and 5.10 {\AA} for YbS are shown in Figs. 3(b) and 4(b), respectively. The calculated band structures show that both YbO and YbS become metallic without a global band gap under high pressure. In contrast to the band structures of YbO and YbS at ambient pressure, the bandwidth of 5\emph{d}-dominating bands increases significantly under high pressure, and band inversion indeed takes place around the X point.  This band inversion will change the parity of occupied bands at the X point.\cite{Fu07} Since the 4\emph{f}-orbital is odd under spatial inversion, while 5\emph{d}-orbital is even, the \emph{d}-\emph{f} hybridization vanishes at the X point, and each band at the X point has definite parity eigenvalues. The parity eigenvalues at all time-reversal invariant momentum points, including one $\Gamma$, three X points and four L points, are listed in Table \ref{parity}. Although both YbO and YbS are metals under high pressure, we still can define the $\mathbb{Z}_2$ topological invariant owing to finite local band gap over the entire BZ.\cite{Fu07} In fact, except the well-known topological metal bismuth, several more topological metals have been predicted, such as the nonmagnetic GdBiPt,\cite{Al10} AuTlTe$_{2}$, and SmS under high pressure.\cite{Yao11, Zhi14}  Since both YbO and YbS have crystal inversion symmetry, the $\mathbb{Z}_2$ invariant can be calculated as the product of half of the parity (Kramers pairs have identical parities) eigenvalues, $\delta_i$, for all the occupied states at the TRIM points, \cite{Fu07}
 \begin{equation}\label{eq1}
    (-1)^{\nu}=\prod_{i=1}^{8}\delta_{i}.
 \end{equation}

 $\nu$=0 reveals trivial band topology in insulating YbO and YbS at ambient pressure, while the nonzero $\nu$=1 in the metallic phases of YbO and YbS reveals the nontrivial band topology of YbO (YbS) with lattice parameter a=4.70 {\AA} (5.10 {\AA}). With a shorter lattice parameter, the band inversion will involve a \emph{d}-band and a lower \emph{f}-band, but the number of band inversion at the X point still is one. The predicted metallic properties of YbO and YbS under high pressure are consistent with experimental results. \cite{werner81,Jarrige13,Leger85} We also note that there is a band crossing between the 5\emph{d}-dominating band and the 4\emph{f}-dominating band. However, the band crossing is absent in compressed SmS and SmO. This band crossing can be explained by the absence of long-range \emph{d}-\emph{f} hybridization. By the extended 3D Bernevig-Hughes-Zhang (BHZ) model on a face-centered cubic (FCC) lattice, the nearest-neighboring (NN) \emph{d}-\emph{f} hybridization allows two gapless nodes in the $\Gamma$(0,0,0)-Z(0,0,2$\pi$) direction.\cite{BHZ}. Because of the shorter lattice parameter of SmS, the next-nearest-neighboring (NNN) \emph{d}-\emph{f} hybridization cannot be ignored, and this NNN hybridization will make the band structure locally gapped in entire BZ. In the YbO and YbS, the NNN \emph{d}-\emph{f} hybridization can be ignored because of the larger crystal lattice.

\begin{figure}[t]
\includegraphics [width=15cm]{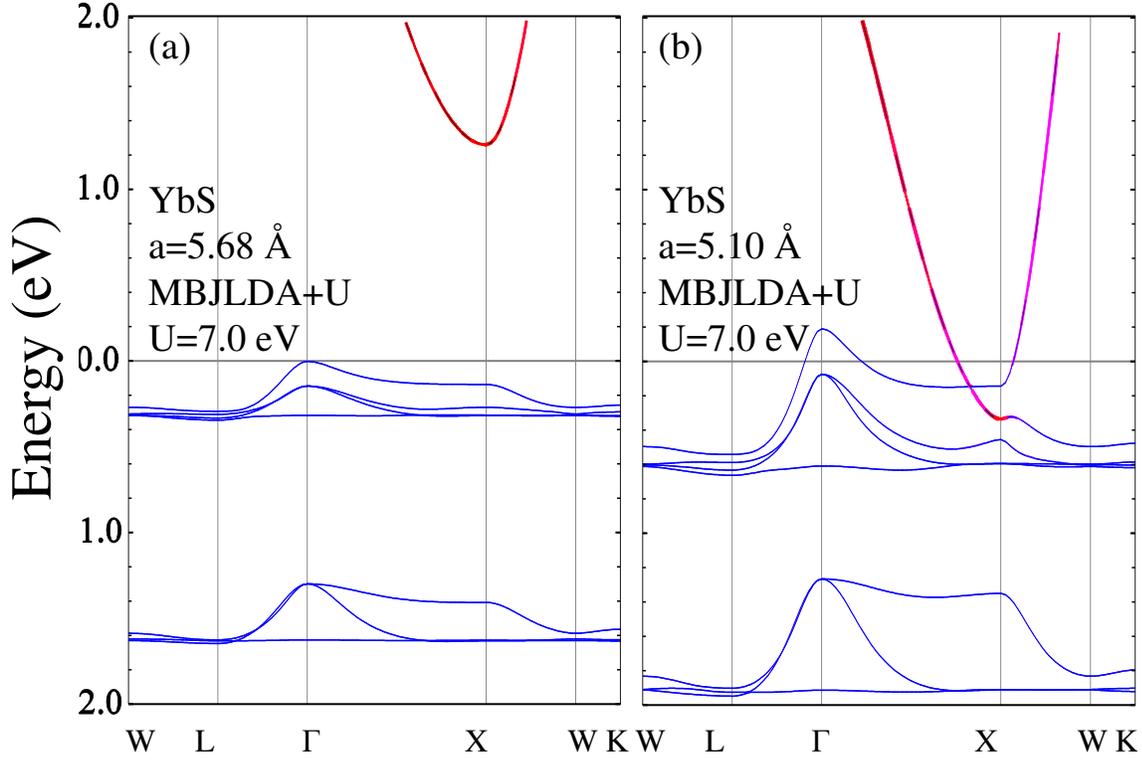}
\caption{\label{fig4} (Color online) Band structure of YbS with experimental lattice parameter a=5.68 (a) and 5.10 {\AA}(b) obtained by GGA+\emph{U} calculation, with \emph{U}=7.0 eV. The weights of 5\emph{d} and 4\emph{f} electron are represented by dark and gray, respectively.}
\end{figure}

\begin{table}
\small
\caption{The product of parity eigenvalues for all occupied states at the $\Gamma$, three X, and four L TRIM points in the BZ, and the
total parity product of all TRIM points. The + and - denote even and odd parity, respectively.}\label{parity}

\begin{tabular}{ccccc}
                              & $\Gamma$    & 3X                         & 4L             & Total  \\
\hline
  insulating YbO     & $-$     & $-$ $-$ $-$       & $-$ ~$-$ ~$-$ ~$-$     & $+$  \\
  insulating YbS     & $-$     & $-$ $-$ $-$       & $-$ ~$-$ ~$-$ ~$-$     & $+$  \\
  metallic YbO       & $-$     & $+$ $+$ $+$       & $-$ ~$-$ ~$-$ ~$-$     & $-$  \\
  metallic YbS       & $-$     & $+$ $+$ $+$       & $-$ ~$-$ ~$-$ ~$-$     &  $-$  \\
\end{tabular}

\end{table}


\begin{figure}[t]
\includegraphics[width=15.0cm]{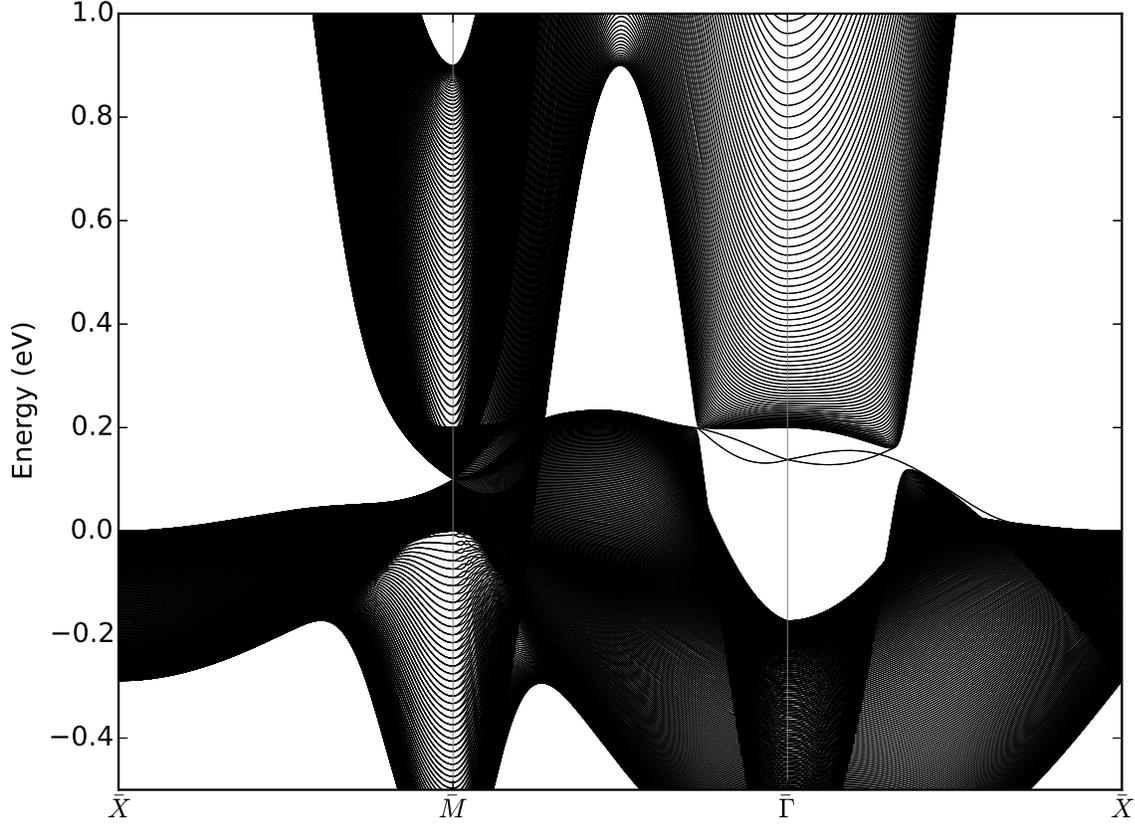}
\caption{\label{fig5}Surface state at (001) surface determined by TB calculation with slab geometry.}
\end{figure}

\textit{Surface state}  By first-principles calculation using maximally localized Wannier functions (MLWFs),\cite{wannier90, Wienwannier} we extract the on-site energy of \emph{f}-orbitals and \emph{d}-orbitals of Yb and the hopping parameters. By tight binding (TB) calculation, we exactly reproduce the band inversion at X point and the band crossing predicted by MBJLDA calculation. The [001] surface state is determined by TB calculation with a slab geometry consisting of 150 layers of YbS unit cells. The surface state is shown in Fig. 5, and the 4\emph{f}-electron pocket is slightly downward shifted to clearly show the surface state. We can find a pair of metallic surface state around the $\bar{\Gamma}$ point. The bulk Dirac cone also is also present near the $\bar{\Gamma}$ point. This surface state is very similar to the [010] surface state of Na$_{3}$Bi,\cite{Na3Bi} in which a pair of metallic surface states coexist with the bulk Dirac cone.

\section{Summary}

In summary, we predict pressure-induced topological phase transition in the chalcogenides YbO and YbS with a rock-salt structure by DFT calculation, despite different magnitudes of correlation for 4\emph{f} electrons in YbO and YbO resulting from a lattice difference. At ambient pressure, because strong correlation localizes the \emph{f}-orbitals, charge transfer between 4\emph{f} and 5\emph{d} orbitals is suppressed, and both YbO and YbS are regular band insulators. However, under high pressure, the bandwidths of 4\emph{f} and 5\emph{d} bands increase, and 5\emph{d} band is below one or several 4\emph{f} bands at the X point with increasing pressure. The 4\emph{f} and 5\emph{d} band inversion of YbO and YbS under pressure change the global band topology and render a topological metallic phase. Although the insulator-metal transition of YbO and YbS under high pressure is well-known, we predict this phase transition is also accompanied by a change in the global band topology. The bulk Dirac cone and a pair of metallic surface states coexist on the [001] surface state.

\begin{acknowledgment}

 J.X. Zhang acknowledges the support from NSF of China (No. 51371010). This research at CSUN was supported by NSF-PREM Grant under No. DMR-1205734. The authors are grateful to N. Kioussis and P. Blaha for valuable discussion.
\end{acknowledgment}

\end{document}